  \documentclass[10pt,prb,nobibnotes,showpacs,showkeys,superscriptaddress,maxnames=6]{revtex4-1}
\usepackage{amsfonts,amsmath,amssymb,graphicx, placeins, bm}
\usepackage{setspace,color,cases}
\usepackage{pstricks}
\usepackage{fdsymbol}

\usepackage{ae,aeguill}
\usepackage[utf8]{inputenc}    
\usepackage[T1]{fontenc}
\newcommand{\ddi}{\hat{\mu}_i\ldotp\hat{\mu}_j -3 (\hat{\mu}_i \ldotp\hat{r}_{ij})(\hat{\mu}_j \ldotp\hat{r}_{ij})}

  \newcommand{\be}{\beta}    \newcommand{\de}{\delta} 
\newcommand{\la}{\lambda} \newcommand{\al}{\alpha}   
  \newcommand{\ep}{\epsilon} \newcommand{\s}{\sigma}
     
 \newcommand{\tend}{\rightarrow}
\newcommand{\equa}[1]{\begin{eqnarray} \label{#1}}
\newcommand{\auqe}{\end{eqnarray}}
\newcommand{\equab}[1]{\begin{widetext}\begin{eqnarray} \label{#1}}
\newcommand{\auqeb}{\end{eqnarray}\end{widetext}}
\newcommand{\tab}[1]{\begin{tabular}{#1}}
\newcommand{\bat}{\end{tabular} \\ }
\begin {document}
\title 
 {Phase diagram of a three-dimensional dipolar model on an fcc lattice.}

\author{V. Russier}
\email[e-mail address: ] {russier@icmpe.cnrs.fr}
\affiliation{ICMPE, UMR 7182 CNRS and UPE 2-8 rue Henri Dunant 94320 Thiais, France.}
\author{Juan J. Alonso}
\email[e-mail address: ] {jjalonso@uma.es}
\affiliation{F\'{\i}sica Aplicada I, Universidad de M\'alaga, 29071 M\'alaga, Spain}
\affiliation{Instituto Carlos I de F\'{\i}sica Te\'orica y Computacional,  Universidad de M\'alaga, 29071 M\' alaga, Spain}
\author{I.~Lisiecki}
\affiliation{Sorbonne Université, CNRS, UMR 8233  De la Molécule aux Nano-Objets: Réactivité, Interactions Spectroscopies, MONARIS, 75005, Paris, France }
\author{A.T.~Ngo}
\affiliation{Sorbonne Université, CNRS, UMR 8233  De la Molécule aux Nano-Objets: Réactivité, Interactions Spectroscopies, MONARIS, 75005, Paris, France }
\author{C.~Salzemann}
\affiliation{Sorbonne Université, CNRS, UMR 8233  De la Molécule aux Nano-Objets: Réactivité, Interactions Spectroscopies, MONARIS, 75005, Paris, France }
\author{S.~Nakamae}
\affiliation{CEA/DRF/IRAMIS/SPEC and CNRS UMR 3680, 91191, Gf sur Yvette, France}
\author{C.~Raepsaet}
\affiliation{CEA/DRF/IRAMIS/SPEC and CNRS UMR 3680, 91191, Gf sur Yvette, France}

\date{\today}   

%
\begin{abstract}
 The magnetic phase diagram at zero external field of an ensemble of dipoles with uniaxial
anisotropy on a FCC lattice is investigated from tempered Monte Carlo simulations.
The uniaxial anisotropy is characterized by a random distribution of easy axes and its
magnitude $\la_u$ is the driving force of disorder and consequently frustration. 
The phase diagram, separating the paramagnetic, ferromagnetic, quasi long range ordered 
ferromagnetic and spin-glass regions is thus considered in the temperature, $\la_u$ plane. 
This system is aimed at modeling the magnetic phase diagram of supracrystals of 
magnetic nanoparticles.
\\~\\ DOI: 10.1103/PhysRevB.102.174410
\end{abstract}
\maketitle
 \section {Introduction}
 \label {intro}
Assembly of magnetic nanoparticles (MNP) in dense packed structures receives a growing 
interest as a way to build new composite materials in a bottom-up strategy. 
Besides promising applications, for instance in nanomedicine~\cite{pankhurst_2009} 
inductor nanotechnology~\cite{garnero_2019},
or mechanical properties~\cite{dreyer_2016,domenech_2019},
the MNP dense packed assemblies present a great fundamental interest both in nanoscale 
magnetism~\cite{bedanta_2009,bedanta_2013}, because of collective effects leading among others to complex magnetic phases, 
and for the knowledge of self organization process. 

Among the diverse dense packed MNP assemblies synthesized experimentally
one can distinguish the random packed ones 
from the long range well ordered 2D or 3D supra-crystals, namely periodic crystals of MNP. 
Examples of the former are the discontinuous metal-insulator multilayers (DMIM) of CoFe in 
alumina matrix~\cite{kleemann_2001a,petracic_2010} 
or pellets made of pressed bare iron oxide MNP reaching the random close packed volume fraction 
limit~\cite{de-toro_2013a,de-toro_2013b}. The latter involve MNP 
assemblies with well controlled size and shape and strongly reduced polydispersity on the first hand 
and the control of the organization, or self-organization process on the other hand.
Slow evaporation and/or solvent destabilization of colloid suspensions, are good examples of 
quite efficient methods to get well ordered supra-crystals. 
Following this route,
ensembles of nearly spherical MNP characterized by a strongly reduced polydispersity and
coated with an organic layer preventing nanoparticle agglomeration self organize according to 
hard sphere like rules in dense packed structures
of either body centered tetragonal (BCT), hexagonal close packed (HCP) or face centered cubic (FCC) 
symmetry~\cite{josten_2017,dreyer_2016,domenech_2019}. 
Similarly, the alternative method based upon a protein crystallization technique, leads to long range 
well organized FCC supra-crystals of iron oxide MNP free of direct contact owing to the protein 
cage~\cite{kasyutich_2008,kasyutich_2010,kostiainen_2011}.
Hence, cobalt or iron oxide MNP long range supra-crystals with FCC or HCP structure
are currently available experimentally.
These supra-crystals quite generally made of nanoparticles coated by a non magnetic layer are
characterized by a resulting volume fraction of the magnetic cores typically in between 0.30 to 0.75
times the maximum value for particles at contact on the corresponding lattice.
In the range of sizes considered, the MNP are single domain and therefore 
bear a large moment referred to as superspin 
(c.a. $2\;10^4$ to $2\;10^5\mu_B$ for iron oxide MNP of 10 to 20 $nm$ in diameter). 
As a result, in these assemblies the mean dipole interaction between MNP taking into account the
volume fraction $E_d/k_B$ reaches up to 100 to 300 K for iron oxide MNP of diameter $\sim{}12\;nm$ 
and up to $3\;10^3$~K for cobalt MNP of diameter $\sim{}10\;nm$.
This makes the low temperature magnetic phase likely to be, at least partly, conditioned by 
dipolar collective behavior, a question which receives a large interest both 
experimentally~\cite{lisiecki_2007a,andersson_2015,mishra_2012} and
theoretically~\cite{petracic_2010}.

 The modeling of the magnetic properties of dense packed structures of MNP is in principle a multi scale 
problem but is fortunately highly simplified for particles whose size falls in the single domain range
if the non collinearity of the surface spins is neglected (the so-called spin canting). Then the 
effective one spin model (EOSM) can be used although this level of approximation has been questioned 
and extended for situations where the core--surface MNP morphology is expected to play a significant 
role~\cite{kachkachi_2006,vasilakaki_2018}. In the framework of the EOSM
and considering the MNP as frozen in position in the system one is led to model an ensemble of MNP 
free of super exchange interaction owing to the non magnetic coating layer and thus interacting only 
through the dipole dipole interaction (DDI) between the moments of the macro spins and undergoing the 
magneto-crystalline anisotropy energy (MAE).
It is well known that strongly coupled dipolar systems present 
a low temperature phase dependent on the underlying structure and the amount of 
disorder~\cite{bouchaud_1993,weis_1993,ayton_1997,weis_2005a,russier_2017} 
which originates either from the structure, the dilution or the MAE through the distribution of easy axes. 
Hence, the so-called super ferromagnetic (SFM) or super spin-glass (SSG) phases can be observed.
Experimental evidence of the SSG state has been given either in randomly distributed MNP or on
well ordered MNP supra-crystals with randomly distributed anisotropy easy axes.
In the case of well ordered supra-crystals for which a FCC lattice is a representative situation, 
the low temperature phase results from the competition between the dipolar induced ferromagnetic
phase of the purely dipolar system (free of MAE), and the frustration introduced by the disorder.
In the case of uniaxial MAE and for a wholly occupied lattice, the latter comes only from either the 
amplitude of the MAE when the easy axes distribution is random, 
or the degree of alignment of the easy axes in the strong MAE limit. 
The former situation corresponds to a  supra-crystal synthesized in the absence of external field. 
This strong MAE limit has been investigated recently in the random
close packed and for MNP on a perfect FCC lattice~\cite{alonso_2019,russier_2020}.

The influence of the random anisotropy such as introduced by the uniaxial MAE with random distribution
of easy axes has already been widely studied in the framework of the random anisotropy model (RAM),
first introduced by Harris {\it et al.}~\cite{harris_1973},
where the spin-spin interaction is quite generally the short-range Heisenberg or Ising exchange one,
the latter being the strong anisotropy limit of the former.
The relevant disorder control parameter is the anisotropy to exchange ratio (D/J) which governs the 
crossover from the FM state (D/J = 0) to the spin-glass state obtained in the $D/J\tend\infty$ 
limit~\cite{chakrabarti_1987}, where $D$ and $J$ are the anisotropy and exchange coupling constants respectively.  
When $D/J$ increases the long-range (LR) FM state first disorders to a quasi long-range order
(QLRO) state with a ferromagnetic character~\cite{itakura_2003}.
The Monte Carlo simulations of Nguyen and Hsiao~\cite{nguyen_2009a,nguyen_2009b} show clearly, 
on the basis of the dynamic behavior, that the PM/FM like magnetic phase transition at weak 
anisotropy disappears with the increase of $D/J$ and a glassy phase transition takes place 
beyond a threshold value. 
This scenario is also in agreement with the simulation results for the spontaneous magnetization in terms
of ($D/J$)~\cite{bondarev_2011}.

In the present work following related contributions on the dipolar Ising 
model~\cite{alonso_2019,russier_2020} devoted to 
the effect of the MAE easy axes texturation in the infinite MAE limit, 
we investigate the magnetic phase diagram from Monte Carlo simulations of 
FCC supra-crystals of MNP interacting {\it via} DDI and undergoing a uniaxial MAE 
whose easy axes are randomly distributed. 
Here the disorder control parameter is the MAE to the DDI strengths ratio as is the case in the RAM.
An important difference with the dipolar Ising model studied in Refs.~\cite{alonso_2019,russier_2020}
is the 3 dimensional nature of the model variables, as is the case in the Heisenberg model with 
important consequences essentially in the spin-glass phase
and with the occurrence of a transverse spin-glass state associated with a QLRO in the FM
region of the phase diagram.
The choice of the FCC structure is mainly dictated by the 
MNP organizations obtained experimentally as outlined above. 
To investigate the phase diagram we calculate the ferromagnetic and overlap spin-glass order parameters, from
which a finite size analysis of the corresponding Binder cumulants, spin-glass and transverse spin-glass 
correlation lengths is performed. We also focus on the finite size behavior of the heat capacity.
The paper is organized as follows. We introduce the model in section~\ref{model}, the simulation details and the observables
we use are then presented in section~\ref{sim_meth} and~\ref{obs} respectively. Section~\ref{result} is devoted to the analysis 
of the results and a conclusion is given in section~\ref{concl}. 

 \section {Model}
 \label   {model}
We model an assembly of MNP free of super exchange interactions, characterized by a uniaxial magneto-crystalline
anisotropy (MAE) and self organized on a supra-crystal of face centered cubic (FCC) structure. 
To this aim we place ourselves in the framework of the effective one spin model where each single domain MNP is assumed to be
uniformly magnetized with a temperature independent saturation magnetization $M_s$.
Hence we consider a system of $N$ dipolar hard spheres of moment 
$\vec{\mu}_i=\mu\hat{\mu}_i$ located on the sites of 
a FCC lattice occupying a total volume $V$, interacting through the usual dipole dipole interaction (DDI) and subjected to a 
one-body anisotropy energy, $-K_{i}v_i(\hat{n}_i\ldotp\hat{\mu}_i)^2$.
$\hat{\mu}_i$ and $\hat{n}_i$ are three dimensional unit vectors and in the following, hatted letters denote unit vectors.
$K_i$, $\hat{n}_i$ and $v_i$ are the anisotropy constant, the easy axis and the volume of the particle $i$ respectively.
We consider a random distribution of easy axes on the unit sphere, namely
the azimuthal angles are randomly chosen while the polar angle distribution follow the probability density $sin(\Theta)/2$.
The MNP ensemble is monodisperse with MNP diameter $d$. 
The particle moment is related to the material saturation magnetization $M_s$ through $\mu=v(d)M_s$ ($v(d)=\pi{}d^{3}/6$).
The Hamiltonian of the system is given by
\equa{beh_1}
  \be H & = & 
        \frac{1}{2} \be \ep_d \sum_{i\neq j} \frac{\ddi}{(r_{ij}/d)^3}
   - \be K v(d) \sum_i (\hat{n}_i\ldotp\hat{\mu}_i)^2
    ~~~ \textrm{with}~ \ep_d = \frac{\mu_0}{4\pi}\frac{\mu^2}{d^3} 
\auqe
where $\hat{r}_{ij}$ is the unit vector carried by the vector joining sites $i$ and $j$,
$r_{ij}$ its length and
$\be=1/(k_BT)$ is the inverse temperature.
The Hamiltonian~(\ref{beh_1}) is the same as that used in~\cite{russier_2017} while in~\cite{alonso_2019},
where the infinite anisotropy limit was considered, the second term of equation~(\ref{beh_1}) was not included 
and the directions of the moments, $\hat{\mu}_i$ were set equal to the easy axes $\hat{n}_i$ representing then 
the Ising axes. Concerning the reduced temperature, instead of the natural choice $Tk_B/\ep_{d}$, we take 
advantage of the $1/r^3$  dependence of the DDI
to introduce the more convenient reduced temperature $T^*=Tk_B/(\ep_{d}(d/d_r)^3)$,
with $d_r=d(\Phi_r/\Phi)^{1/3}$
where $\Phi$ and $\Phi_r$ are respectively the volume fraction, $(N/V)\pi{}d^3/6$, and a reference value 
(here, the maximum value for hard spheres on a FCC lattice).
This choice of $T^*$ equivalent to measuring the temperature in terms of a dipole dipole energy 
weighted by the volume fraction instead of its maximum value at contact~\cite{note_t*,russier_2020}.
Equation~(\ref{beh_1}) is then rewritten as
\begin{subequations}\label{beh_2}
\equa{beh_2a}
  \be H  &=&  \frac{1}{T^*} \left( \frac{1}{2} \sum_{i\neq j} \frac{\ddi}{(r_{ij}/d_r)^3} 
              - \la_u \sum_i (\hat{n}_i\ldotp\hat{\mu}_i)^2 \right )
   ~~~~ \la_u = \frac{Kv(d)}{\ep_d} (d_r/d)^3
\auqe
\equa{beh_2b}
    \equiv \frac{1}{T^*} \left( \frac{1}{2} \sum_{i\neq j} 
                        \hat{\mu}_i \bar{T}_{ij}  \hat{\mu}_j 
                        - \la_u \sum_i (\hat{n}_i\ldotp\hat{\mu}_i)^2 \right )
\auqe
\end {subequations}
which introduces the MAE coupling constant $\la_u$.

The simulation box is a cube with edge length $L_s=\sqrt{2}Ld_r$ and the total number of dipoles is $N=4L^3$.
(In the following without loss of generality we consider the case of a FCC lattice with $d_r=d$).
We consider periodic boundary conditions by repeating the simulation cubic box identically in the 3 dimensions.
The long range DDI interaction is treated through the Ewald summation technique~\cite{allen_1987,weis_1993}, 
with a cut-off $k_c=10\;k_m$, $k_m=(2\pi/L_s)$, in the sum of reciprocal space and the $\al$ parameter of the 
direct sum chosen is $\al=5.80$, a value which permits to limit the sum in direct space to the first image 
term~\cite{kusalik_1990,weis_1993}. The Ewald sums are performed with the so-called 
conductive external conditions~\cite{allen_1987,weis_1993}, i.e. the system is embedded in a medium with 
infinite permeability, $\mu_s=\infty$, which is a way to avoid the demagnetizing effect and thus 
to simulate the intrinsic bulk material properties regardless of the external surface and system 
shape effects. 
For the different values of $\la_u$ considered in the following,
the simulations are performed for system sizes up to either $L=7$ ($N=1372$) or $L=8$ ($N=2048$).
\subsection {Simulation method}
\label{sim_meth}

In order to thermalize in an efficient way our system presenting strongly frustrated states,
we use parallel tempering algorithm~\cite{hukushima_1996,earl_2005} (also called tempered Monte Carlo) for our 
Monte Carlo simulations. Such a scheme is widely used in similar systems~\cite{alonso_2010,alonso_2019,russier_2020}, 
and we do not provide the details. The method is based on the simultaneous simulation runs of 
identical replica for a set of temperatures $\{T^*_n\}$ with exchange trials of the configurations 
pertaining to different temperatures each $N_M$ Metropolis steps according
to an exchange rule satisfying the detailed balance condition. The set of temperatures is chosen
in such a way that it brackets the transition temperature
while ensuring a satisfying rate of exchange between adjacent temperature configurations.
Our set $\{T^*_n\}$ is either an arithmetic distribution or an optimized one in order to make the 
exchange rate between adjacent paths as constant as possible in the whole range of $\{T^*_n\}$ 
according to the efficient constant entropy increase method~\cite{sabo_2008}.
In the present work we take $N_M=10$, the number of temperatures is in between 36 and 48 
according to the value of $L$ and the amplitude of temperatures in the set $\{T^*_n\}$ 
for $L\le{}7$ and up to 96 for $L=8$. 
When necessary, precise interpolation for temperatures between the points actually 
simulated are done through reweighting methods~\cite{ferrenberg_1988}.
We use $t_0$ Monte Carlo steps (MCS) for the thermalization and  the averaging is
performed over the $(t_0,2t_0)$ interval following MCS with $t_0=5\;10^5$.

We deal with frozen disorder situations where each realization of the easy axes distribution 
$\{\hat{n}_i\}$ defines a sample. Accordingly, a double averaging process is performed first relative 
to the thermal activation, {\it i.e.} the Monte Carlo step, and second on the whole set of $N_s$ samples.
Consequently, the mean value of an observable $A$, results from a double averaging denoted in the following as
$[<A>]$ where $<.>$ corresponds to the thermal average on the MC sampling for a fixed realization of
the axes distribution and $[.]$ to the average over the set of samples considered.
The number of samples necessary to get an accurate result depends strongly on the value of $\la_u$.
Obviously, for $\la_u=0$, $N_s=1$ should be sufficient for a very long MC run in order to get a satisfying average. 
In practice, we use $N_s=10$ for $\la_u=0$. We find that $N_s$ of the order of 100 to 200 is sufficient up to $\la_u\leq{}1.5$ 
whereas accurate results for $\la_u=4$ require the use of at least $N_s=400$ realizations. 
However, it is worth mentioning that accurate results for
the heat capacity $C_v$ are much less demanding than for the overlap order parameter and about 50 to 100 realizations allow to 
get very well averaged $C_v$ up to $\la_u=100$. In any case, in the present work we limit ourselves to $\la_u=4$ in the  
computations of 
the spin-glass overlap order parameter $q_2$ and related properties.
The error bars of the averaged quantities are deduced from the mean squared deviations of the sample to sample fluctuations.
The simulation code is massively parallelized, all temperatures and the 2 replicas running together, and the typical
CPU time is {\it c.a.} 40h for the complete run of one sample of $L=7$ ($N=1372$) on the Intel Xeon E5 processors
at the CINES center. 

\subsection {Observables}
\label {obs}

Our main purpose is the determination of the transition temperature between the paramagnetic and the ordered 
phase and on the nature of the latter, namely ferromagnetic or spin-glass, in terms of $\la_u$. 
For the PM/FM transition, 
we consider the spontaneous magnetization 
\equa{m_1}
   m = \frac{1}{N} \left\Arrowvert  \sum_i \hat{\mu}_i \right\Arrowvert
\auqe
computing its moments, $m_n=[<m^n>]$, n = 1,2 and 4.
We compute also the nematic order parameter $P_2$ together with the instantaneous nematic direction, $\hat{d}$ which are 
the largest eigenvalue and the corresponding eigenvector respectively of the tensor
$  
 \bar{Q} = \frac{1}{N} \sum_i (3\hat{\mu}_i\hat{\mu}_i - \bar{I})/2
$~~\cite{weis_1993}, where $\bar{I}$ is the unit tensor of cartesian components $I_{\al\be}=\de_{\al\be}$. 
The spontaneous magnetization can also be studied in the ordered phase from the mean value
projected total magnetization on the nematic direction, which defines 
\equa{m_la}
   m_{d} =  \frac{1}{N} \sum_i \hat{\mu}_i.\hat{d} 
\auqe 
We compute the mean value $m_{d1}=[<|m_{d}|>]$ and the moments $m_{dn}=[<m_{d}^n>]$, with $n=2,4$.
To locate the transition temperature, $T^*_c$, as usually done, we will use the finite size scaling (FSS) analysis of the Binder
cumulant~\cite{binder_1981} which is defined either from the moments $m_n$ or  $m_{dn}$ characterized by 3 or one degree of freedom respectively
 \equa{bm_la}
   B_m = \frac{1}{2} \left( 5 - 3 \frac{m_{4}}{m_2^2} \right) , 
   ~~~~~~~~~~~~~~~ B_{md} = \frac{1}{2} \left( 3 - \frac{m_{d4}}{m_{d2}^2} \right)
 \auqe
From these normalizations, $B_m,B_{md}\tend{}1$ in the long range FM phase and $B_m,B_{md}\tend{}0$ in the limit $L\tend\infty$ in the disordered 
PM phase. For the PM/SG transition, we consider the usual overlap order parameter~\cite{fernandez_la_2009a,viet_2009b}
\equa{q}
  q^2 = \sum_{\al\be} \lvert q_{\al\be} \lvert ^2~, ~~~~ \textrm{with} ~~~~ q_{\al\be} = \frac{1}{N} \sum_i \mu_{i\al}^{(1)}\mu_{i\be}^{(2)}~,
\auqe
where the superscripts $^{(1)}$ and $^{(2)}$ stand for two independent replicas of an identical sample
and $\al, \be$ denote the cartesian coordinates. From $q^2$ 
we calculate the mean value $q_2=[<|q^2|>]$ and $q_4 = [<q^4>]$ and the corresponding Binder cumulant,
 \equa{bq}
    B_q = \frac{1}{2}(11 - 9 \frac{q_4}{q_2^2})
 \auqe 
 In order to determine the PM/SG transition temperature it may be more convenient to use the spin-glass correlation length, usually defined 
from a fit of the spin-glass overlap parameter correlation function on its Ornstein-Zernike form, assuming that it follows from
a so called $\Phi^4$ theory~\cite{cooper_1982,ballesteros_2000}
\equa{xi_sg}
  \xi = \frac{1}{2 sin(k_m/2)} \left( \frac{q_2(0))}{q_2(\vec{k}_m)} -1 \right)^{1/2} ~~
  \textrm{with}  ~~~ q_2(\vec{k}) = \left[\left< \sum_{\al,\be} q_{\al\be}(\vec{k}) q^*_{\al\be}(\vec{k}) \right>\right]
   ~~~~ q_{\al\be}(\vec{k}) =  \frac{1}{N} \sum_i \mu_{i\al}^{(1)}\mu_{i\be}^{(2)} e^{i\vec{k}\vec{r}_i} ~ ,
\auqe
with $\vec{k}_m=(k_m,0,0)$.
By exploiting the non dimensional character of the ratio $\xi/L$ one is led to deduce $T^*_c$ as the crossing point of the
$\xi/L$ curves in terms of $T^*$ for different values of $L$.
Finally, the magnetic susceptibility, $\chi{}_m$ and the heat capacity , $C_v$ are calculated from the magnetization and the energy 
fluctuations respectively
\equa{fluct}
  \chi_m = \frac{N}{T^*}\left[ \left(<m^2> - <m>^2 \right) \right] ~~~, ~~ C_v = \frac{1}{NT^{*2}} \left[ \left( <H^2> - <H>^2 \right) \right]
\auqe
%

 \section {Results}
 \label {result}
\begin {figure}[h!]
  \includegraphics [width = 0.75\textwidth , angle =  00.00]{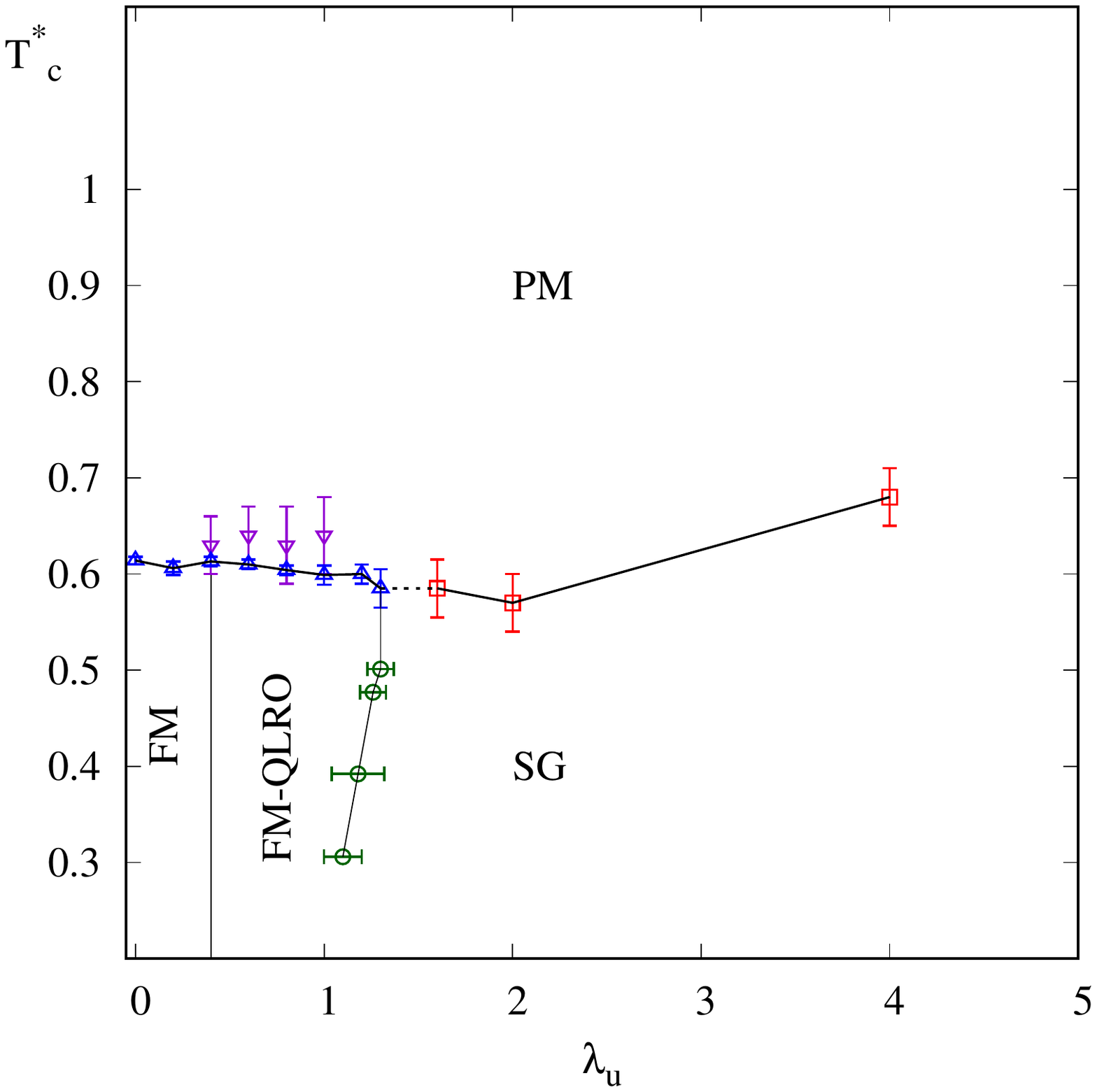}
  \vskip -0.05\textheight
  \caption {\label {diag_phase_rand}  
  Phase diagram in the $(T^*,\la_u)$ plane separating the PM, FM, SG regions. We also indicate the
  estimated range in $\la_u$ corresponding to the FM-QLRO and transverse spin-glass phase. 
  Open upward triangles: PM/FM line; open squares: PM/SG line; 
  open downward triangles : transverse spin-glass transition temperature, $T_{xy}$ ;
  open circles: FM/SG line obtained from the 
  behavior of $B_m$ in terms of $L$ at constant $T^*$.
     }
  \end {figure}

The phase diagram, which summarizes the present work, is displayed in figure~(\ref{diag_phase_rand}).
It separates the paramagnetic (PM), ferromagnetic (FM) and the spin-glass (SG) phases.
The important feature of $T^*_c(\la_u)$ along both the PM/FM and the PM/SG lines is its very weak dependency 
with $\la_u$, the overall variation being limited to $0.6\leq{}T^*_c\leq{}1.0$.

\begin {figure}[h!]
  \includegraphics [width = 0.90\textwidth , angle =  00.00]{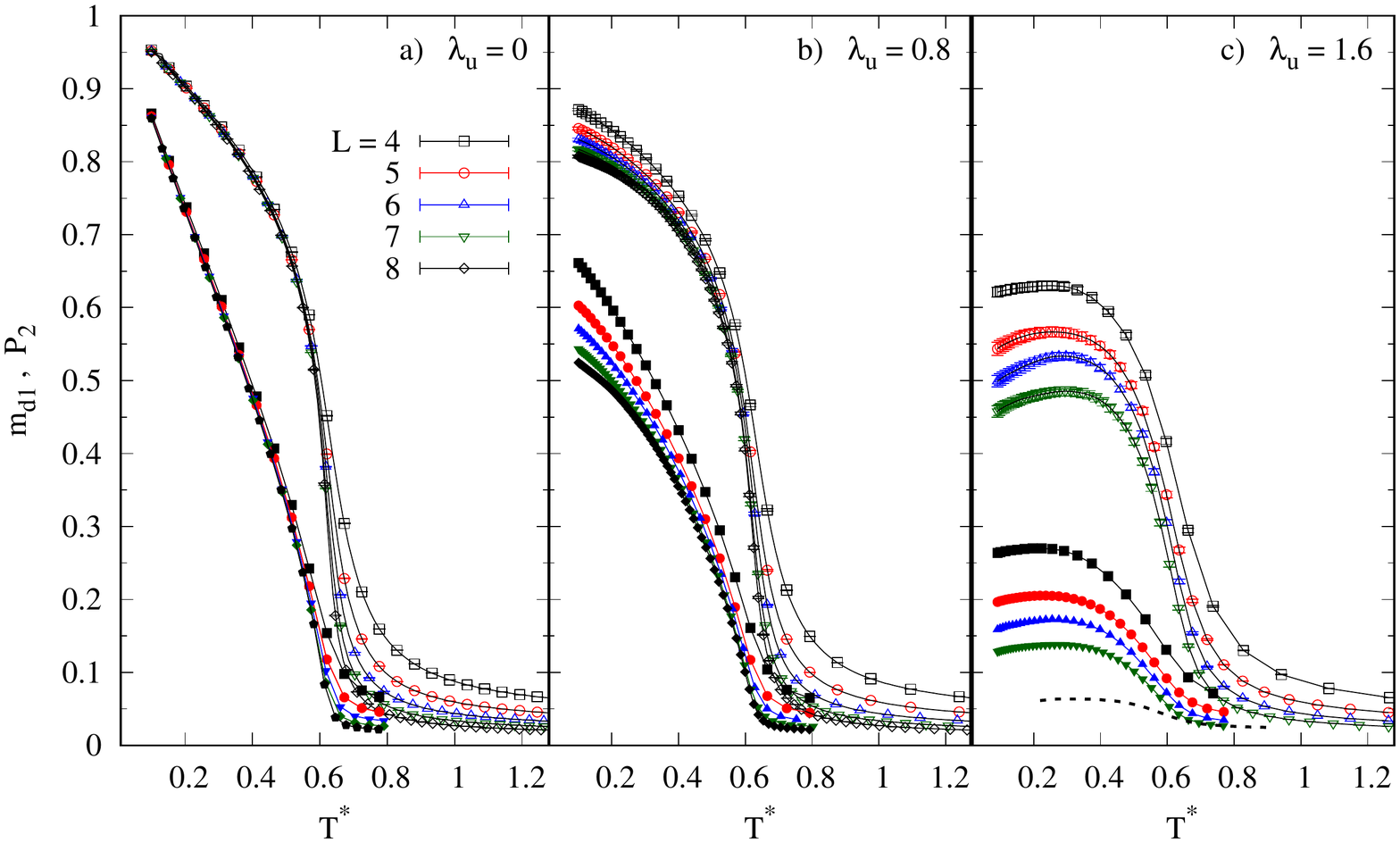}
  \vskip -0.05\textheight 
  \caption {\label {m_p2_rand_0-2}
  Polarization $m_{d1}$, open symbols and nematic order parameter, $[<P_2>]$ solid symbols in 
  terms of $T^*$ for a) $\la_u=0$ , b) $\la_u=0.8$ and c) $\la_u=1.6$ for different system sizes 
  $L=4$ to 8 ($N=256$ to 2048) for $\la_u=0$ and 0.8, and $L=4$ to 7 for $\la_u=1.6$, as indicated.
  In c) the dashed line represents $P_2$ for $\la_u=2$ and $L=7$ to illustrate the gradual vanishing 
  of the nematic order at low $T^*$ with the increase of $\la_u$.
  The lines result from the interpolation performed by the reweighting 
  method~\cite{ferrenberg_1988}. 
     }
  \end {figure}

\begin {figure}[h!]
  \includegraphics [width = 0.95\textwidth , angle = -00.00]{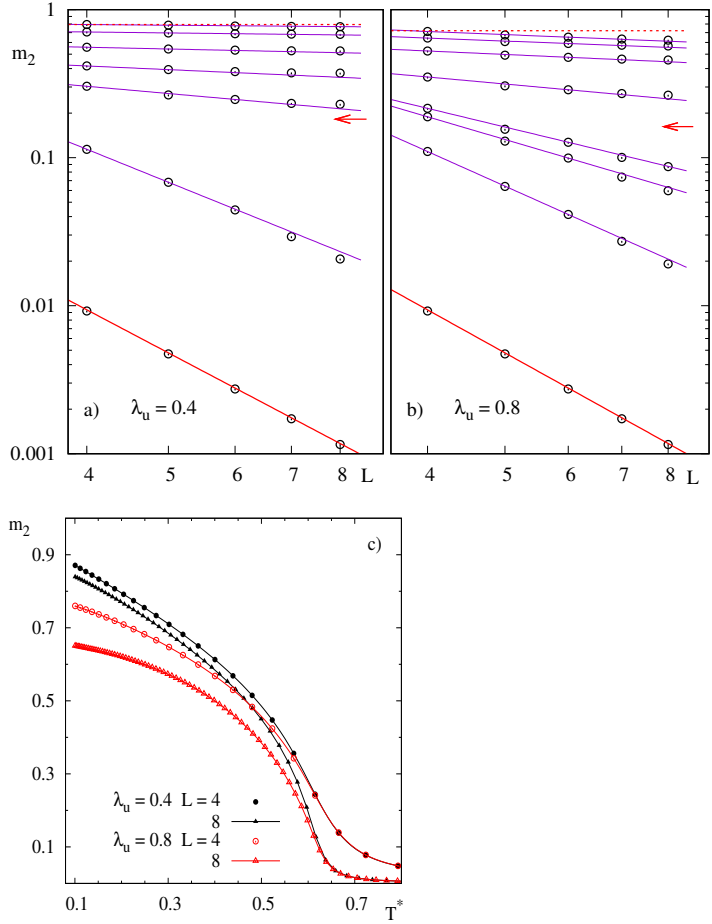}
  \vskip -0.12\textheight
  \caption {\label {lnm2_rand_04-08}
  Log-log plot of $m_2$ in terms of $L$ for~: a) $\la_u=0.4$ and 
  $T^*=0.2028$, 0.3050, 0.4481, 0.5402, 0.5913, 0.6833, 1.910 from top to bottom; 
  b) $\la_u=0.8$ and
  $T^*=0.1983$, 0.3084, 0.4429, 0.5651, 0.6263, 0.6385, 0.6874, 1.9100.
  The dotted lines are guides to the eyes for $m_2=cst.$ at the smallest temperature; 
  the solid lines are linear interpolation determined from $L=4$ and $L=6$. 
  excepted at bottom where they indicates
  the $1/L^3$ behavior in the PM phase. The arrow localizes $T^*_c$.
  c) $m_2$ in terms of $T^*$ for $\la_u=0.4$ at $L=4$ and 8 ($N$ = 256 and 2048) compared to the case $\la_u=0.8$ 
  which shows the onset at $\la_u=0.4$ of the decrease of $m_2$ with respect to $L$ at low temperature. 
  Error bars in all panels are smaller than the size of the symbols.
       }
  \end {figure}

 Before going in the details of its determination, a qualitative overview of the evolution with respect to the MAE 
coupling constant $\la_u$ combined with the finite size effect of both the ferromagnetic and nematic order parameters, 
namely the magnetization $m$ and the eigenvalue $P_2$ is useful. 
This is displayed in figure~(\ref{m_p2_rand_0-2}). 
%
The rising of the nematic order parameter below a threshold temperature, close to the transition 
temperature $T^*_c$ is indicative of the occurrence of a collective spontaneous direction in the system 
induced by the DDI which competes with the random anisotropy contribution.
It is worth mentioning that the finite value of the order parameters in the PM phase, well above $T^*_c$, 
results from the the short range correlations in finite size systems~\cite{itakura_2003} and is a quite general 
result~\cite{ayton_1997,weis_2005a,klopper_2006,alkadour_2017}.  
The curves of figure~(\ref{m_p2_rand_0-2}) suggest clearly an evolution of the low temperature phase
from a well ordered FM phase in figure~(\ref{m_p2_rand_0-2}\;a) corresponding to the pure dipolar case, $\la_u=0$ 
where the PM/FM transition is well established~\cite{bouchaud_1993,weis_1993,russier_2017,alkadour_2017} 
to a gradual crossover to the absence of FM order, figure~(\ref{m_p2_rand_0-2}\;c) with the complete 
disappearance of the latter at $\la_u=2$.
Between these two regimes, figure~(\ref{m_p2_rand_0-2}\;b), 
the similarity of $m$ and $P_2$ in terms of $T^*$ with the pure dipolar case
remains only on a qualitative level 
since then both order parameters $P_2$ and $m$ decrease with the system size $L$ conversely to
the case at $\la_u=0$.  
The behavior of both $P_2$ and $m$ at $\la_u=0.2$ is characterized, as is the case at $\la_u=0$ by
the merging at low $T^*$ of the curves corresponding to different values of $L$, indicating a finite
value of $P_2$ and $m$ in the thermodynamic limit ($L\tend\infty$) 
and we find the onset of the decrease with $L$ of the ferromagnetic order parameters at 
$\la_u\simeq{}0.4$, as is clarified in figure~(\ref{lnm2_rand_04-08}c) 
where we compare the behavior of $m_2(T^*)$ with the system size between $\la_u=0.4$ and 0.8. 
%
 \FloatBarrier
  \subsection {Ferromagnetic transition}
  \label {pm_fm_line}
\begin {figure}[h!]
  \hskip -0.10\textwidth
  \includegraphics [width = 1.15\textwidth , height = 0.75\textheight , angle =  00.00]{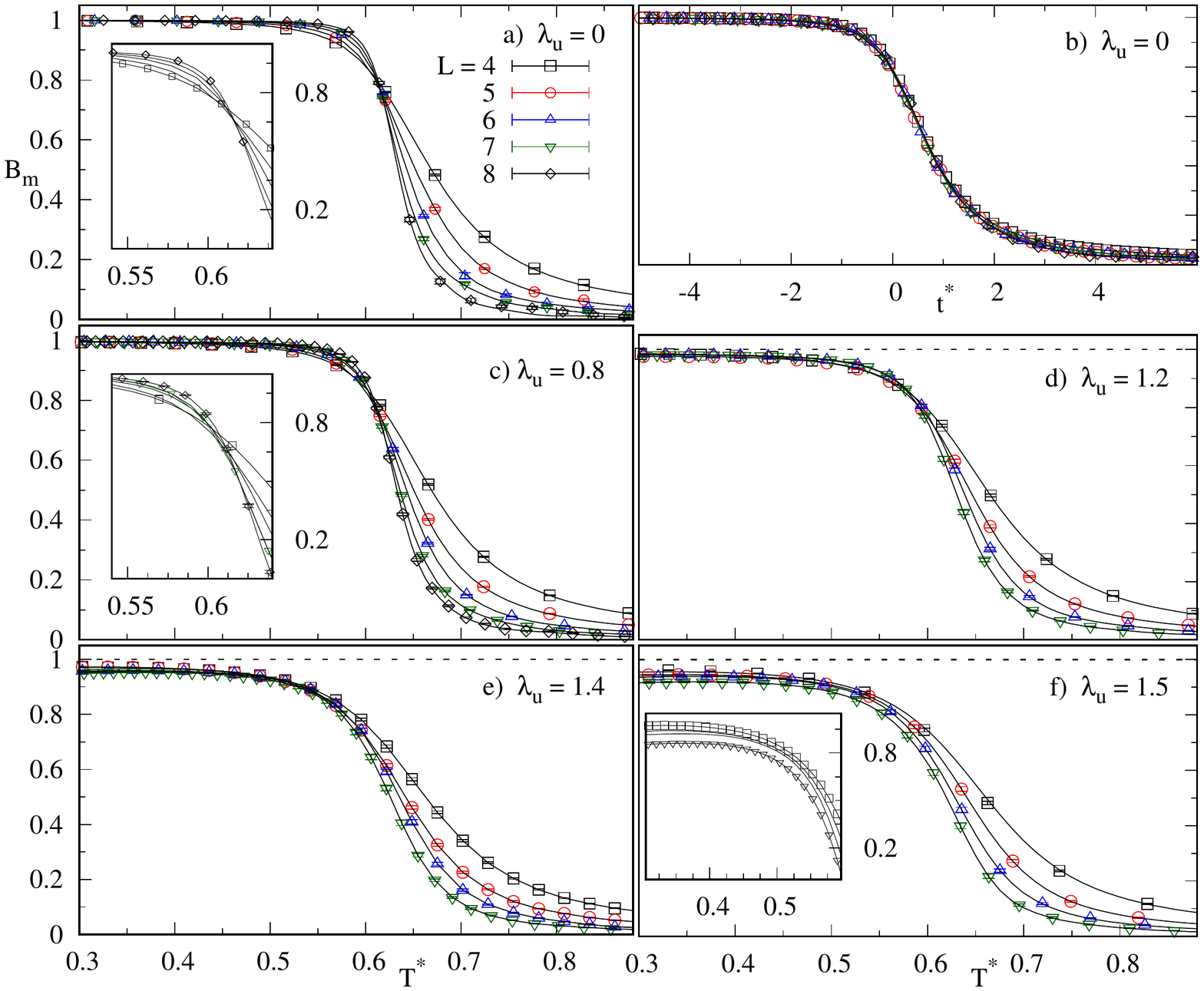}
  \vskip -0.05\textheight
  \caption {\label {bm_rand_0-15}
  Magnetization Binder cumulant for different ssytem sizes as indicated. 
  In the pure dipolar case (a) and b)). a) in terms of $T^*$; b) rescaled according to equation~(\ref{scal_1})
  with $t^*=(T^*-T_c^*)L^{1s/\nu}$, $T_c^*=0.614$ and $\nu=0.692$~\cite{bruce_1974}.
  For different values of $\la_u$, as indicated (c) to f)); 
  including the vicinity, $\la_u=1.4$, e) and f) beyond, $\la_u=1.5$, the onset of the SG phase of the magnetic phase diagram.
  The lines are the interpolations according to the reweighting
  method~\cite{ferrenberg_1988}. The dotted line in d) to f) is a guide to the $B_m=1$ limiting value,
  which is no longer reached for $\la_u\ge{}1.2$.
     }
  \end {figure}

\begin {figure}[h!]
  \includegraphics [width = 0.95\textwidth , angle =  00.00]{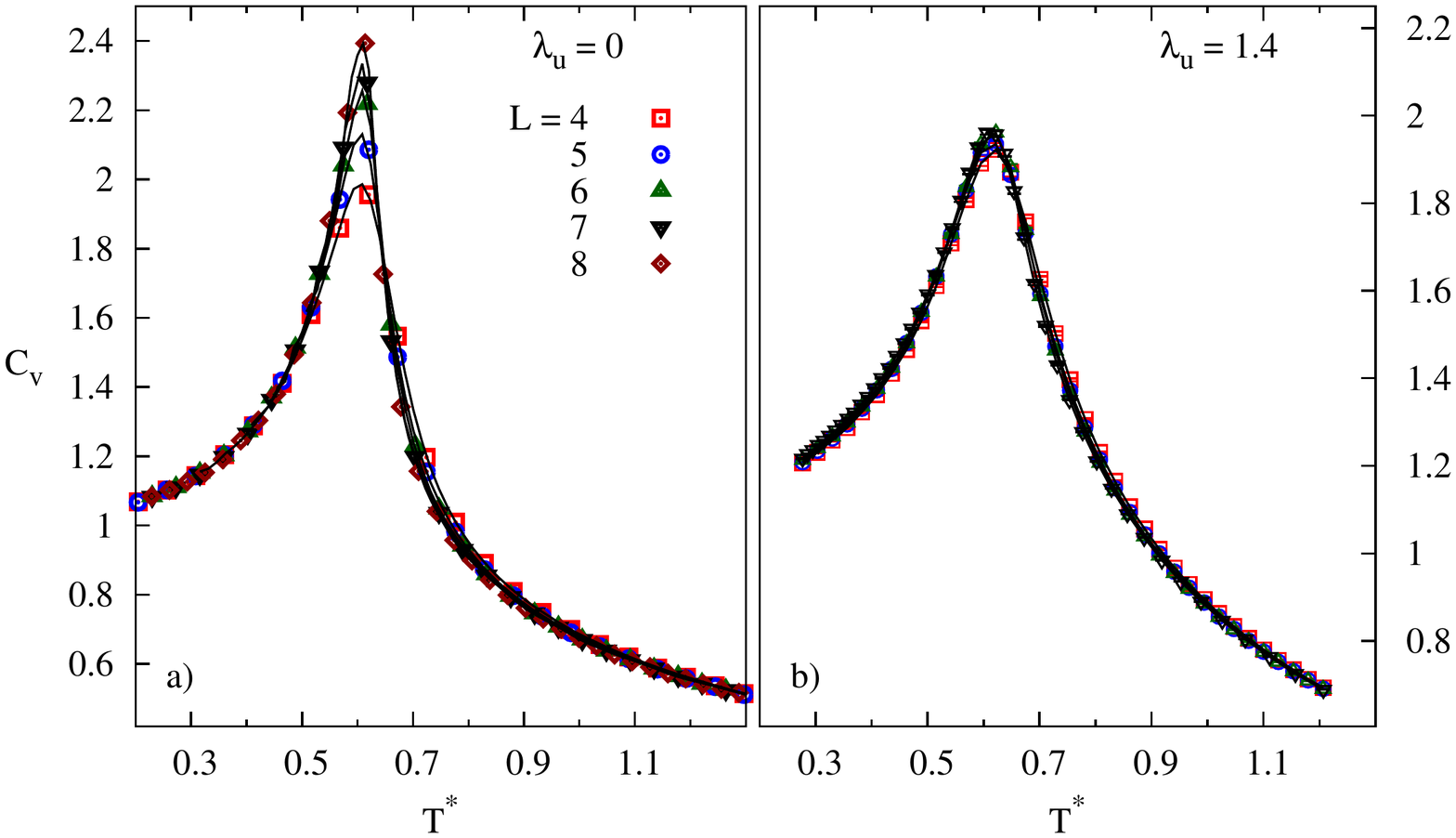}
  \vskip -0.05\textheight
  \caption {\label {cv_rand_0-14}
   Evolution of the heat capacity $C_v$ with the system size $L$ for a):~the pure dipolar case ($\la_u$ = 0)
   and $L=4$ to 8 and b):~$\la_u=1.4$ and $L=4$ to 7.
   The error bars are smaller than the size of the symbols.
   The lines result from the interpolation performed by the reweighting method~\cite{ferrenberg_1988}.
     }
  \end {figure}

For small values of $\la_u$, starting from $\la_u=0$, we check the ferromagnetic nature of the low temperature phase 
and determine the transition temperature $T^*_c$ by following the FSS and thus we look for the crossing point of the 
Binder cumulant, $B_m(L)$ curves for different values of $L$, consequence of the scaling behavior 
\equa{scal_1}
  B_m = b_m((T^*-T^*_c)L^{1/\nu})~.
\auqe
  However, the system sizes considered in the present work are too small to determine accurate values of the exponent $\nu$.
  At $\la_u=0$ we get $T^*_c=0.614\pm{}0.004$ in agreement with 
Refs.~\cite{bouchaud_1993,alkadour_2017,russier_2017}. The spontaneous magnetization orientates along the <111> directions 
in agreement with results of the literature~\cite{bouchaud_1993,alkadour_2017} and 
as Alkadour {\it et al.}.~\cite{alkadour_2017} we do not find any spin reorientation at low temperature.
Moreover we have checked that equation~(\ref{scal_1}) is very well reproduced using $\nu=0.692$~\cite{bruce_1974},
the value of $\nu$ obtained for dipoles on lattice of cubic symmetry from renormalization group developments,
(see figure~(\ref{bm_rand_0-15}\;a,b)). 
%
Another important feature of the PM/FM transition is the strong finite size dependence of both the heat capacity $C_v$
and the magnetic susceptibility $\chi_m$ 
with a peak located at $T_L^*(A)$, ($A=C_v,\chi_m$),  and $T_L^*(A)\tend{}T^*_c$ in the $L\tend\infty$ 
limit~\cite{papakonstantinou_2015,russier_2020}. This behavior, leading to a divergence of $\chi_m$ (not shown) at $T_c$ 
is displayed on figure~(\ref{cv_rand_0-14}a) for $C_v$~\cite{rem_cv}. 

\begin {figure}
  \includegraphics [width = 0.90\textwidth , angle =  00.00]{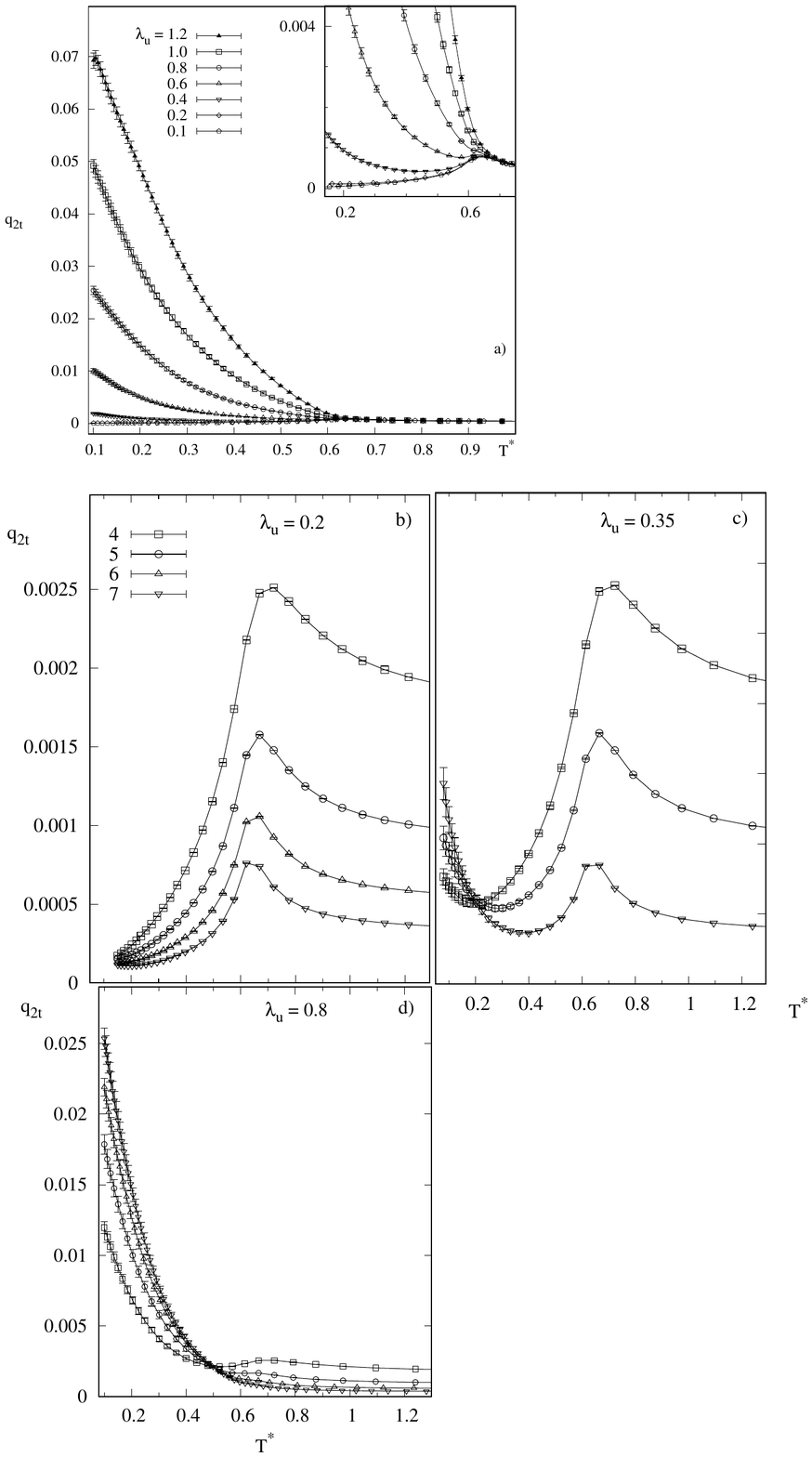}
 \vskip -0.05\textheight
 \caption {\label {q2t_01-12} 
  Spin-glass transverse overlap order parameter defined in the plane normal to the nematic direction for 
  a) $\la_u=0.1$ to 1.2 as indicated and $L=7$. Inset: detail of the hump in the vicinity of the PM/FM transition temperature. $q_{2t}$ for
  $\la_u=0.1$ and 0.2 are only slightly distinguishable at the scale of the inset. \\
   Finite size effect on $q_{2t}$, for $L=4$ to 7 as indicated and b) $\la_u=0.2$, c) $\la_u=0.35$ showing
   the very beginning of the low temperature rising up in the latter case. 
   d) Case $\la_u=0.8$, well inside the QLRO region of the phase diagram.
    }
  \end {figure}

\begin {figure}
  \includegraphics [width = 0.77\textwidth , angle =  00.00]{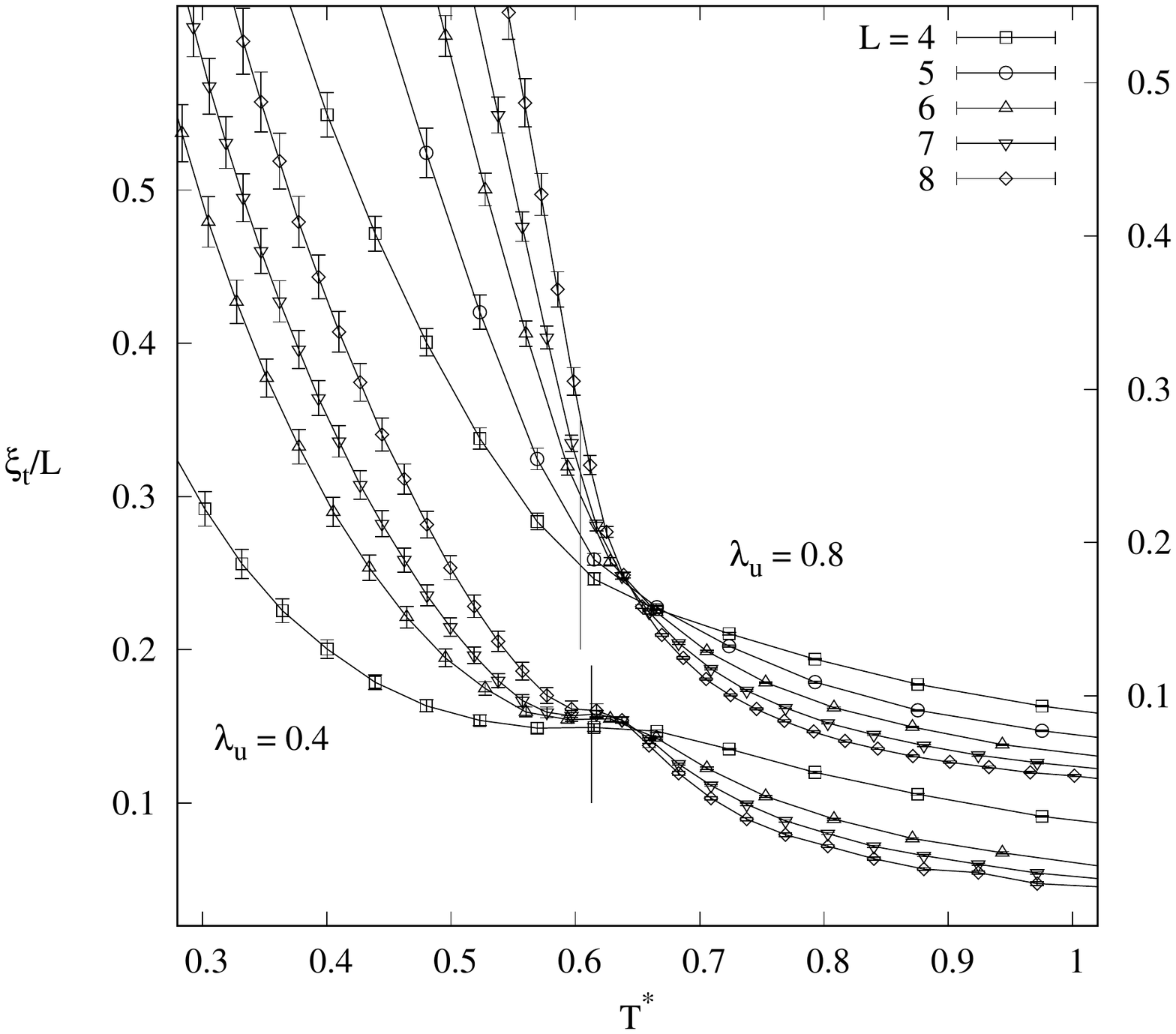}
  \vskip -0.05\textheight
  \caption {\label {xit_04-08}
  Transverse spin-glass correlation length reduced by $L$ for $\la_u=0.4$, bottom and 0.8, top, showing the crossing
  point behavior in terms of $T^*$ for different values of $L$.
  The vertical lines visualize the value of the PM/FM transition temperature $T^*_c$ obtained from $B_m$.
    }
  \end {figure}

When $\la_u$ increases in the low MAE regime, the behavior of $m_2$ with respect to the system size changes as we observe 
a decrease of the magnetization with the increase of $L$ at low temperature (see figure~(\ref{m_p2_rand_0-2}\;b)).
Although we cannot rule out a crossover for much larger values of $L$ leading to a convergence of $m(T^*)$ at fixed 
and low $T^*$ we interpret this behavior as a FM state with quasi long range order
with $m(L\tend\infty)=0$ well below $T^*_c$, instead of a long range FM one.
In the absence of MAE, the different $m(L,T^*)$ merge at low $T^*$ (see figure~(\ref{m_p2_rand_0-2}\;a) 
and Ref.~\cite{russier_2017}) 
which is no longer the case beyond $\la_u\sim{}0.4$ as can be seen in figures~(\ref{m_p2_rand_0-2}\;b) 
and~(\ref{lnm2_rand_04-08}\;c).  
 %
This FM-QLRO is observed beyond $\la_u\simeq{}0.4$ as shown in figure~(\ref{lnm2_rand_04-08}) where $m_2$ in terms of $L$ 
in log scale (a and b) and $m_2$ in terms of $T^*$ for different values of $L$ (c) are displayed for $\la_u=0.4$ and 0.8. 
In any case the paramagnetic behavior, namely $m_2\sim{}1/N$, is reached sufficiently far from $T^*_c$.
The FM-QLRO is related to the onset of the transverse spin-glass state~\cite{beath_2007} where the components of the moments normal to the
nematic direction freeze in a spin-glass like state and corresponds to the mixed state of the phase diagram outlined by 
Ayton~{\it et al.}~\cite{ayton_1997}. 
The transverse spin-glass sate is defined in the FM region of the phase diagram and characterized by the transverse
overlap order parameter, $q_{2t}$, obtained by using the transverse components 
$\hat{\mu}_{it}=(\hat{\mu}_i-(\hat{d}\ldotp\hat{\mu}_i)\hat{d})/\left\Arrowvert(\hat{\mu}_i-(\hat{d}\ldotp\hat{\mu}_i)\hat{d})\right\Arrowvert$
of the moments instead of $\hat{\mu}_i$ in equation~(\ref{q}).
From the behavior of $q_{2t}$ in terms of $T^*$, given in figure~(\ref{q2t_01-12}\;a) for increasing values of $\la_u$,   
we see that no transverse spin-glass state is expected for $\la_u<0.4$. Indeed, 
$q_{2t}$ not only remains very 
small on the whole range of $T^*$ but presents no noticeable rising up at low temperature and is moreover a decreasing 
function of $L$ down to very low temperature up to $\la_u={0.35}$ as shown in figure~(\ref{q2t_01-12}\;b,c) where we also 
compare figure~(\ref{q2t_01-12}\;d)
to the case $\la_u=0.8$ well inside the FM-QLRO region of the phase diagram.
 %
We emphasize that this is related to the onset at $\la_u=0.4$ of the the $L$--dependence of $m_2(T^*)$ at low $T^*$ 
as mentioned above.
Given the very small values taken by $q_{2t}$ for $\la_u=0.35$, we cannot interpret the crossing point of the $q_{2t}(T^*)$
curves for different values of $L$ as the onset of the transverse spin-glass state.
We emphasize that $q_{2t}(T^*)$ when $T^*<0.6$ remains smaller or comparable to its value 
at the hump ($\sqrt{q_{2t}}\le{}0.04\ll{}1$). This is no longer the case for instance at $\la_u=0.8$, figure~(\ref{q2t_01-12}a,\;d), 
in the FM-QLRO region of the phase diagram.
The corresponding transition temperature, $T^*_{xy}$ 
below which the FM and transverse spin-glass orders coexist is determined
by the crossing point of the $\xi_{t}/L$ curves where $\xi_{t}$ is the transverse spin-glass correlation length
as obtained from equation~(\ref{xi_sg}) with the overlap order parameter replaced by its transverse equivalent.
The result we get is that $T^*_{xy}$ is very close to and seems slightly larger than the PM/FM transition temperature $T_c^*$, 
see figures~(\ref{xit_04-08}) and~(\ref{diag_phase_rand}). 
 %
The small difference between $T^*_{xy}$ and $T_c^*$ is however not really 
meaningful given the error bars (see figure~(\ref{diag_phase_rand})).
Therefore we are led to conclude from the present simulations that $T^*_{xy}$ and $T_c^*$ are likely to coincide.

\begin {figure}
  \includegraphics [width = 0.55\textwidth , angle =  00.00]{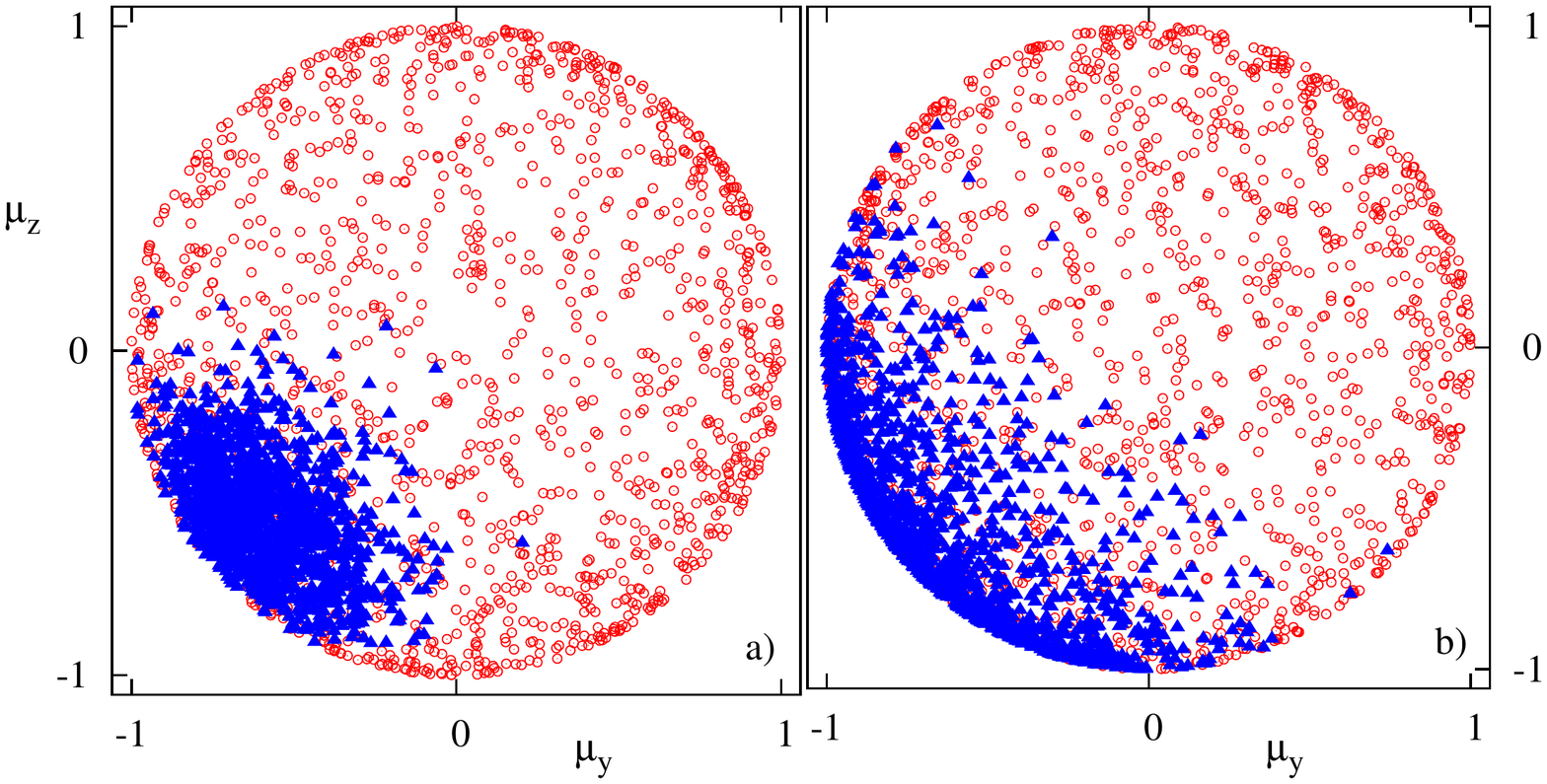}
  \vskip -0.05\textheight \hskip -0.05\textwidth
  \caption {\label {conf_rand_0-08}
  Examples of the instantaneous moments distribution $(\mu_y,~\mu_z)$ at $T^*=0.1$, solid triangles and in the PM phase, open circles
  for $\la_u=0$, a) and 0.8, b). 
           }
  \end {figure}

  Increasing $\la_u$, we still find a PM/FM transition up to $\la_u=1.3$ as indicated by the behavior of $B_m$ as shown in
figure~(\ref{bm_rand_0-15}\;c,d). 
Indeed, we still get a crossing point in temperature, 
allowing the determination of $T_c^*$.
Moreover, the orientation of the magnetization in the FM phase remains very close to the <111> 
directions (see figure~(\ref{conf_rand_0-08}b)). 
Of course the distribution of the moments around the mean polarization direction depends
largely on $\la_u$ and this can be easily visualized from an instantaneous moment configuration as is done in 
figure~(\ref{conf_rand_0-08}\;a,b) and estimated from the variation of the nematic order parameter mean value, 
$P_2$ in terms of $\la_u$ at constant temperature. 
Using a simple model to represent the distribution of the moment around the nematic direction, 
as for instance 
\makebox{$P(\Theta)\propto{}sin(\Theta)[exp(-\Theta^2/(2\s^2))]$}, 
we can get the variance $\s$ 
of this distribution for a given value of $P_2$. For instance at $T^*=0.1$, we get 
$\s=0.226$, 0.458 and 0.656 for $\la_u=0$, 0.8 and 1.2 where $P_2=0.859$, 0.543 and 0.304 respectively.
%
  %
Then, at $\la_u=1.2$, the value taken by $B_m$ at low temperature is slightly smaller 
than 1, and the clear crossing observed for $T^*<T^*_c$ at lower values of $\la_u$ becomes merely a merging
at least for sufficiently large values of $L$. 
At $\la_u= 1.3$ the different $B_m(L)$ curves merge in a very limited range of temperature, and 
from $\la_u=1.4$ and beyond $B_m(L)$ is a decreasing function of $L$ whatever the value of $T^*$ indicating 
the absence of PM/FM transition as displayed in 
figure~(\ref{bm_rand_0-15}\;e,f).
This is an indication of the change in the nature of the transition from the PM/FM to the PM/SG one
or at least to a PM/glassy phase.
This is corroborated by the reduction of the finite size dependence of the heat capacity, as can be seen on 
figure~(\ref{cv_rand_0-14}b), since no singular finite size behavior of $C_v$ is expected in a PM/SG 
transition~\cite{ogielski_1985}. This latter feature is more evident on the dipolar component of $C_v$ in the 
strong MAE coupling case (discussed in more detail in section~\ref{inf_lam_u} and in figure~(\ref{c_an_dip})).
We conclude that the PM/FM line extends from $\la_u=0$ to $\la_u^{(s)}=1.3$ and we
expect the low temperature phase to present a SG character beyond this value. 

\FloatBarrier
\subsection {Spin-glass transition}
\label {pm_dg_line}
\begin {figure}
  \includegraphics [width = 0.75\textwidth , angle =  00.00]{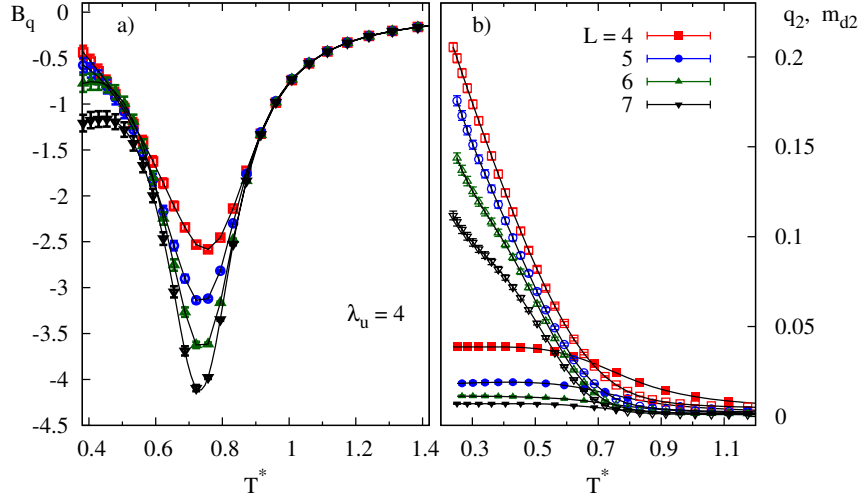}
  \vskip -0.05\textheight 
  \caption {\label {bqsg_rand_4}
  Spin glass Binder cumulant $B_q$, a) and 
  second moment of the spin-glass overlap order parameter, $q_2$, open symbols, and of the magnetization, $m_{d2}$, 
  solid symbols, b), for $\la_u=4$ in terms of the system size $L$.
     }
  \end {figure}

\begin {figure}
  \includegraphics [width = 0.75\textwidth , angle =  00.00]{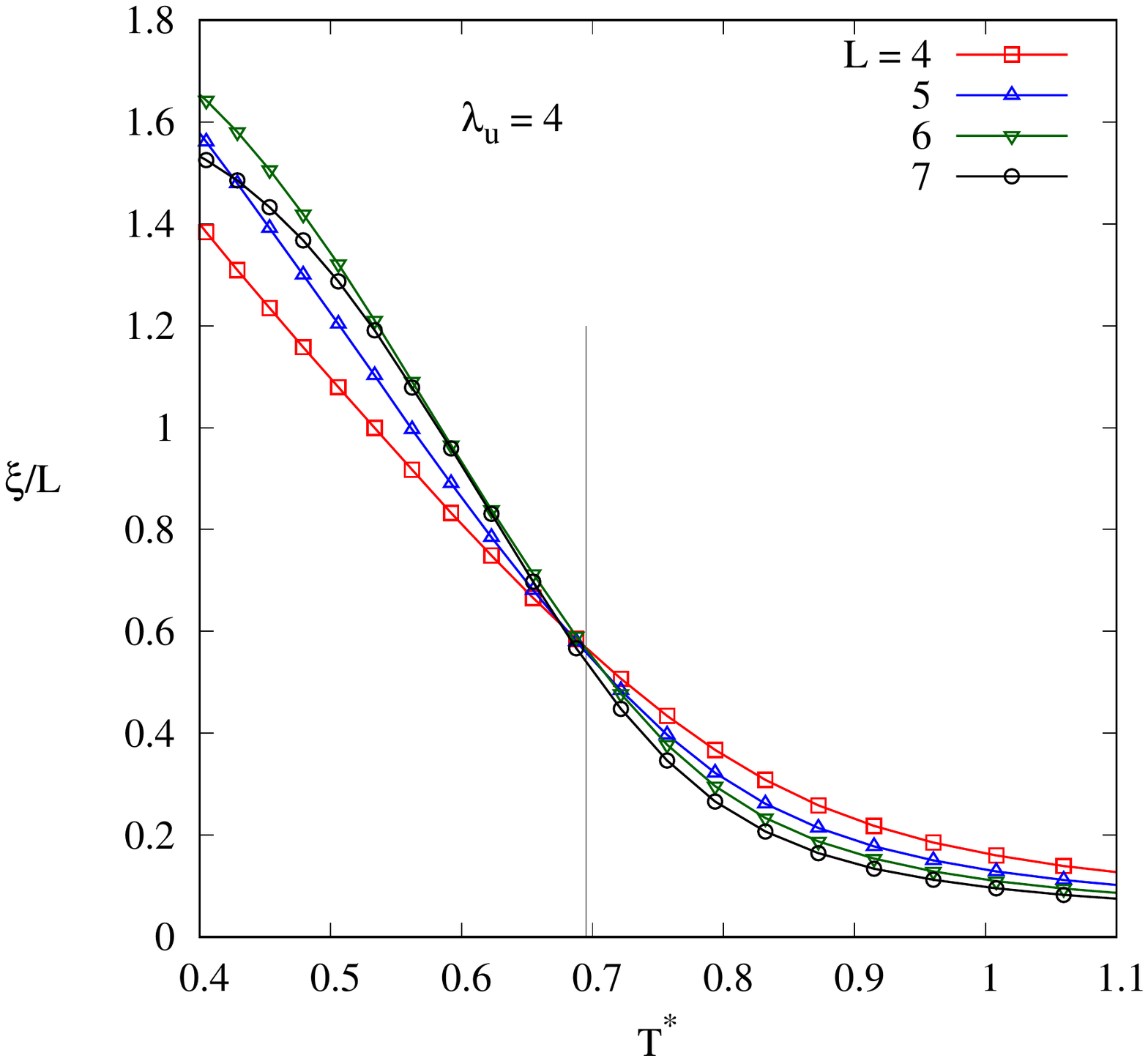}
  \vskip -0.05\textheight
  \caption {\label {xi_sg_rand_4}
  Spin glass correlation length reduced by the system size in terms of $T^*$ for $\la_u=4$. 
  The vertical line indicates the value of $T^*_c$ obtained from the interpolation at $1/L=0$
  of the linear fit in terms of $1/L$ of the Binder cumulant dip temperature.
     }
  \end {figure}

Beyond $\la_u=1.4$ we are clearly in the PM/SG transition region. 
As is the case for the PM~/~chiral glass transition of 
the 3D Heisenberg spin-glass model the Binder cumulant~(\ref{bq})
does not present a well defined crossing point as in the FM/PM transition.
It is instead characterized by a dip in the $B_q$ negative region, which deepens with the increase of $L$. 
The location of this dip must converge towards $T^*_c$~\cite{viet_2009a,viet_2009b,ogawa_2020}. 
Our simulations confirm such a behavior for $B_q$ 
as can be seen in figure~(\ref{bqsg_rand_4}) where $B_q$ is displayed for $\la_u=4$.
 %
Therefore, for $\la_u>1.5$, we determine the value of $T^*_c$ as the crossing point of
the $\xi/L$ curves for different values of $L$ on the one hand and the interpolation to $1/L=0$
of the linear fit in $1/L$ of the value of $T^*$ at the dip in $B_q$ on the other hand. 
Because of the heaviness of the computations, we consider only
$\la_u=1.6$, 2 and 4. The two above mentioned determinations of $T^*_c$ are in quite good agreement,
as can be seen in figure~(\ref{xi_sg_rand_4}) corresponding to $\la_u=4$.
 %
Furthermore the value of $T^*_c$ in the limiting case, $\la_u\tend\infty$, 
which is the so-called random axes dipoles (RAD) model~\cite{fernandez_2009,alonso_2017}
is known from the literature, $T^*_c(\infty)=0.95\pm0.05$~\cite{russier_2020}.

The FM/SG transition line below $T^*_c$ is not easy to locate. 
We have nevertheless estimated $\la_{u}(T^*_c)$ along the FM/SG line from the condition that $B_m(L)$
at $T^*$ constant is an increasing (decreasing) function of the system size $L$ in the FM (SG) side of 
this line, with however a rather large uncertainty (see figure~(\ref{diag_phase_rand})).

\FloatBarrier
\subsection {Large $\la_u$ limit}
\label {inf_lam_u}
\begin {figure}[h]
  \includegraphics [width = 1.00\textwidth , angle = 00.00]{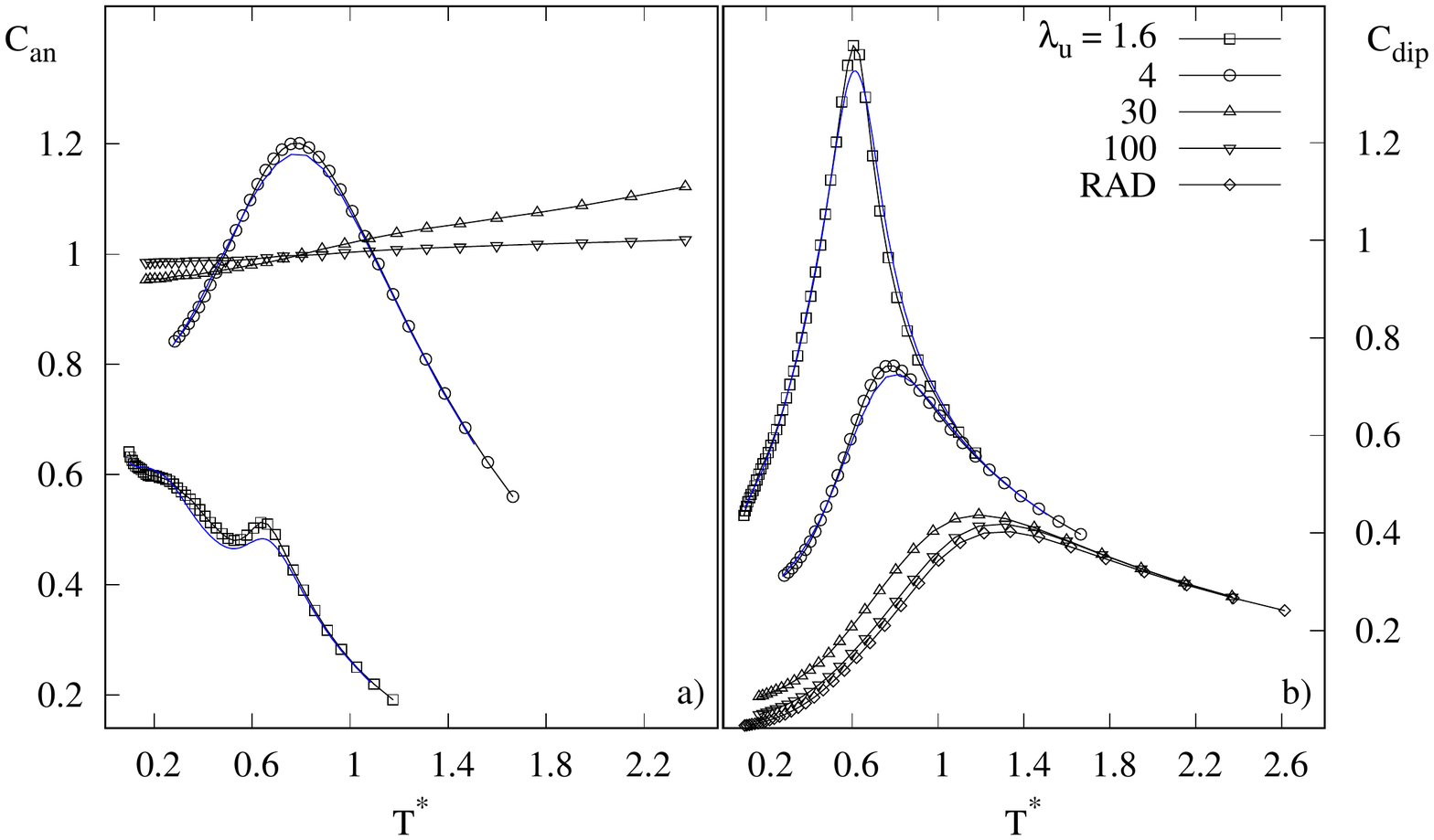}
  \vskip -0.05\textheight
  \caption {\label {c_an_dip}
  Anisotropy, a) and dipolar, b) components of the heat capacity. In the RAD case the total capacity includes only the
  $C_{dip}$ component.
  Symbol: $L=7$ for $\la_u\leq{}4$ and $L=4$ otherwise. Solid blue lines for $\la_u\leq{}4$: $L=4$.
     }
  \end {figure}

In the strong $\la_u$ limit, the MAE contribution to the total Hamiltonian~(\ref{beh_1}) tends to align
the moments $\hat{\mu}_i$ on the anisotropy axes $\hat{n}_i$ making the model coincide with the dipolar
Ising model~\cite{klopper_2006,fernandez_2009,alonso_2010,alonso_2017}, 
the anisotropy axes playing the role of the Ising axes. In this limit as is the the case for the 
$D/J\tend\infty$ of the Heisenberg RAM~\cite{chakrabarti_1987}, one is left with a much simpler system
since the continuous 3D unit vector per site transforms in the two valued scalar variable of the Ising model.
Therefore, it is important to determine the value of $\la_u$ beyond which the system behaves as a dipolar 
Ising system. 
To do this we note 
that when $\la_u\gg{}1$ the moments directions $\hat{\mu}_i$ get close to the axes $\hat{n}_i$ leading to
$\hat{\mu}_i\simeq{}s_{i}\hat{n}_i$ with $s_i=\pm1$ and thus the Hamiltonian from equation~(\ref{beh_2b}) can be written as 
\equa {beh_4}
  \be H = \frac{1}{T^*}
     \left( (1/2) \sum_{i\neq{}j} s_i J_{ij} s_j - \la_u \sum_i (\hat{\mu}_i.\hat{n}_i)^2 \right)
\auqe
Where $J_{ij}$ are the coupling constants $\hat{n}_i \bar{T}_{ij} \hat{n}_j$, independent of the
configuration of the moments $\{\hat{\mu}_i\}$ in the framework of the frozen easy axes distribution.
 Equation~(\ref{beh_4}) reflects the fact that in this limit the MAE depends on the fluctuations 
 $((\hat{\mu}_i.\hat{n}_i)^2 -1)$
 conversely to the DDI which depends on the configuration of the $\{s_i\}$.
We then have two independent sets of variables per site namely $\{s_i\}$ and $(\hat{\mu}_i.\hat{n}_i)^2$.
The second one, describing the MAE are totally uncoupled and represent two degrees of freedom per site
($\hat{\mu}_i$ are 3-dimensional unit vectors) and 
thus contribute, according to the Dulong and Petit law, 
or equivalently to the equipartition of energy, as $Nk_B$ 
to the total heat capacity, or in other words  ($C=C_v/(Nk_B)$)
\equa {c_an}
C_{an}(T) = \frac{1}{Nk_B} \frac{d [<E_{an}>]}{d T} \tend 1 
\auqe
Then equation~(\ref{c_an}) is a clear criterion to determine whether the system of interacting dipoles plus MAE is Ising like
since it is an indication that the variables $\{s_i\}$ and $(\hat{\mu}_i.\hat{n}_i)^2$ become uncoupled and equation~(\ref{beh_4})
holds. In such a case, we expect the dipolar component of $C$, $C_{dip}$ to be close to the $C$ of the corresponding dipolar 
Ising model which does not include the component $C_{an}$, as for instance in the RAD case.
From the heat capacity curves displayed in figure~(\ref{c_an_dip}) we conclude that the Ising dipolar behavior is reached 
beyond $\la_u=30$. It is worth mentioning that a similar conclusion was obtained in the case of the totally textured 
distribution of easy axes~\cite{russier_2020}. 
 %
\FloatBarrier
\section  {Conclusion}
 \label   {concl}
 In this work, we have determined a significant part of the magnetic phase diagram of an ensemble of dipoles with uniaxial anisotropy
located on the nodes of a FCC lattice from tempered Monte Carlo simulations.
This is motivated first by the search for the conditions under which a super-ferromagnetic phase induced by DDI 
can be reached in the supra-crystals of MNP synthesized experimentally,
and more generally by the determination of the nature of the ordered low temperature phase in these systems. 
The nature of the low temperature ordered phase is found successively FM and SG with increasing the MAE strength.
From the behavior of $m_2$ with respect to $L$ at low $T^*$ beyond $\la_u=0.4$ we interpret the FM phase
in this region of the phase diagram as a FM-QLRO phase.
Moreover the FM-QLRO phase is related to a transverse spin-glass state.

The crucial points to relate our findings to actual experimental situations is to determine the corresponding
value of the $\la_u$ parameter and the reference temperature
$T_r$ defined as $T^*=T/T_r$ (see section~\ref{model}). $\la_u$ and $T_r$ are related to the physical characteristics of the MNP,
the uniaxial anisotropy constant $K$ and the saturation magnetization $M_s$ by
$\la_u=(24/\mu_0)(K/M_s^2)(\Phi_r/\Phi)$ and $T_r=\mu_0/(4\pi{}k_B)(\pi/6\;M_s)^2d^3(\Phi/\Phi_r)$
respectively from the definitions given after equations~(\ref{beh_1},\ref{beh_2a}). 
As typical examples,
in the case of maghemite or cobalt MNP we get a lower bound $\la_u>5$ which means that in the absence of texturation of
the easy axes, the corresponding supra-crystals are in the SG region of the phase diagram at low temperature.
A precise determination of $T_r$ and thus of the transition temperature $T_c=T^*_{c}T_r$ 
necessitates the precise knowledge of both the MNP size $d$ and $M_s$ which because of finite size and surface chemistry effects
deviates from its bulk value, making thus $T_r$ specific to a given system, and not only to a given material. 
One may also wonder on the effect of the temperature dependence of the relevant experimental parameters, $M_s$ and
$K$ on the validity of such a model. Concerning $M_s$ one has to keep in mind that the ordering temperature
(either the Néel or the Curie temperature) of the usual MNP materials ($c.a.~\sim$ 850\;-\;1000~K) is much larger than the 
room temperature which is somewhat an upper bond for the MNP properties of interest and thus only a very small
variation of $M_s$ ($\sim{}M_s(0)(1-(T/T_c))^b,~b\sim3/2$) is expected. 
Concerning $K$, since it is in general an effective anisotropy constant including different contributions 
(magnetocrystalline, shape, surface chemistry), a precise evaluation of $K(T)$ is not a simple task.
However, our results show only a very weak dependence of the transition temperature with respect to $\la_u$.
On an other hand an important bound is the
$T_b/T_r$ ratio, where $T_b$ is the blocking temperature related to the MAE through $T_b=Kv(d)/(ck_B)$, 
the constant $c$ being specific to the measurement considered ($c\simeq{}30$ for magnetometry).
Below $T_b$ the moments are blocked in the MAE potential well during the measurement and thus the DDI induced 
transition can be observed only if $T_b<T_{r}T^*_c$.
From the definition of $T_b$ we easily get $T_b/T_r=\la_u/c$ and consequently the DDI induced transition will be
experimentally observable only if $\la_u<cT^*_c$, the sufficient condition being in any case $\la_u<c/2$ 
according to our results.

Finally, we also have obtained the limiting value beyond which the DDI plus uniaxial MAE model is assimilable to a 
1D dipolar Ising model. This value , $\la_u\simeq{}30$, seems larger that the experimental expectation of 
$\la_u$. As a result, modeling the MNP ensemble by a dipolar Ising model may be not strictly speaking justified.
%
  \section  {Acknowledgements.}
  \label    {acknow}
  This work was granted an access to the HPC resources of CINES under the allocations 2018-A0040906180 
  and 2019-A0060906180 made by GENCI, CINES, France.  
  We thank the SCBI at University of Málaga and IC1 at University of Granada for generous allocations of computer time. 
  J.-J.~A. thanks for financial support from grant FIS2017-84256-P (FEDER funds) from the Agencia Espa\~nola de Investigaci\'{o}n.
  \FloatBarrier
  \newpage
 
   \FloatBarrier

\begin{thebibliography}{56}
\bibitem [{not()}]{note_t*}%
  \BibitemOpen
  \href@noop {} {\ }\bibinfo {note} {Introducing $d_r=d(\Phi_r/\Phi)^{1/3}$
  involving the volume fraction $\Phi$ of the system ($(N/V)\pi{}d^3/6$) and a
  reference value, $\Phi_r$ (for instance the maximum value for hard spheres on
  a FCC lattice), $(1/N)\sum_{i\ne{}j}^{N}(1/(r_{ij}/d_r)^3)=f(N)$ is found
  nearly independent of both the underlying structure and $\phi$. As a result
  $(1/N)\sum_{i\ne{}j}^{N}([\hat{\mu}_i\ldotp\hat{\mu}_j-3(\hat{\mu}_i\ldotp\hat{r}_{ij})(\hat{\mu}_j\ldotp\hat{r}_{ij})]/(r_{ij}/d)^3)$
  in equation~(\ref{beh_1}) is rewritten as
  ~$(\Phi/\Phi_r)(1/N)\sum_{i\ne{}j}^{N}([\hat{\mu}_i\ldotp\hat{\mu}_j-3(\hat{\mu}_i\ldotp\hat{r}_{ij})(\hat{\mu}_j\ldotp\hat{r}_{ij})]/(r_{ij}/d_r)^3)$.
  Accordingly, $((\ep_d/k_B)(\Phi/\Phi_r))$ is a convenient measure of the
  temperature for this system from which our definition of $T^*$,
  $T^*=k_{B}T/(\ep_d(\Phi/\Phi_r))$ follows.}\BibitemShut {Stop}%
\bibitem [{rem()}]{rem_cv}%
  \BibitemOpen
  \href@noop {} {\ }\bibinfo {note} {We do not expect $C_v$ to diverge at the
  transition because of the scaling relation, $\al=2-3\nu$ in 3D and according
  to ($C_v^{*}(L)\sim{}C_v^{*}(\infty)-bL^{\al/\nu}$) since in any case we do
  not expect $\nu<2/3$ in our system.}\BibitemShut {Stop}%
\end{thebibliography}
%
\end   {document}